# Two-dimensional covalent crystals by chemical conversion of thin van der Waals materials

Vishnu Sreepal[1,2], Mehmet Yagmurcukardes[3], Kalangi S. Vasu[1,2], Daniel J. Kelly[4], Sarah F. R. Taylor[2], Vasyl G. Kravets[5], Zakhar Kudrynskyi[6], Zakhar D. Kovalyuk[7], Amalia Patanè[6], Alexander N. Grigorenko[5], Sarah J. Haigh[1,4], Christopher Hardacre[2], Laurence Eaves[5,6], Hasan Sahin[8], Andre K. Geim[5], Francois M. Peeters[3], Rahul R. Nair*[1,2]

[1]National Graphene Institute, University of Manchester, Manchester, M13 9PL, UK.

[2]School of Chemical Engineering and Analytical Science, University of Manchester, Manchester, M13 9PL, UK.

[3]Department of Physics, University of Antwerpen, Groenenborgerlaan 171, B-2020 Antwerpen, Belgium.

[4]School of Materials, University of Manchester, Manchester, M13 9PL, UK.

[5]School of Physics and Astronomy, University of Manchester, Manchester, M13 9PL, UK.

[6] School of Physics and Astronomy, University of Nottingham, Nottingham, NG7 2RD, UK.

[7]Institute for Problems of Materials Science, The National Academy of Sciences of Ukraine, Chernivtsi Branch, Chernivtsi 58001, Ukraine.

[8]Department of Photonics, Izmir Institute of Technology, 35430, Izmir, Turkey.

**ABSTRACT:** Most of the studied two-dimensional (2D) materials have been obtained by exfoliation of van der Waals crystals. Recently, there has been growing interest in fabricating synthetic 2D crystals which have no layered bulk analogues. These efforts have been focused mainly on the surface growth of molecules in high vacuum. Here, we report an approach to making 2D crystals of covalent solids by chemical conversion of van der Waals layers. As an example, we use 2D indium selenide (InSe) obtained by exfoliation and converted it by direct fluorination into indium fluoride ($InF_3$), which has a non-layered, rhombohedral structure and therefore cannot be possibly obtained by exfoliation. The conversion of InSe into $InF_3$ is found to be feasible for thicknesses down to three layers of InSe, and the obtained stable $InF_3$ layers are doped with selenium. We study this new 2D material by optical, electron transport **and Raman measurements and show that it is a semiconductor with a direct bandgap of 2.2 eV, exhibiting high optical transparency across the visible and infrared spectral ranges. We also demonstrate the scalability of our approach by chemical conversion of large-area, thin InSe laminates obtained by liquid exfoliation into $InF_3$ films. The concept of chemical conversion of cleavable thin van der Waals crystals into covalently-bonded non-cleavable ones opens exciting prospects for synthesizing a wide variety of novel atomically thin covalent crystals.**

Chemical modification of materials has proved to be a powerful tool for obtaining novel materials with desired and often unusual properties[1-6]. Following the exfoliation of graphene[7], the family of two-dimensional (2D) materials was populated either by direct exfoliation of layered bulk crystals[7-10] or by epitaxial growth techniques[11-14]. Also, the concept of using an existing 2D material as an atomic scaffold for synthesizing novel 2D materials has been demonstrated by hydrogenated and fluorinated graphene, called graphane[5] and fluorographene[6], respectively. Similarly, ion exchange was used to modify existing 2D materials and synthesize $MoS_2$, $WS_2$, $Cu_2SnS_3$, $ZnS$, $PbS$, etc. However, these techniques are usually limited to producing 2D layers of the known layered crystals[15-17]. In contrast to layered crystals, where atomic layers are held together by weak van der Waals forces and hence can be separated by "brute" mechanical action, covalent solids cannot be exfoliated. One can also imagine the chemical conversion of two or more atomic layers of a van der Waals solid into a 2D covalent material of controlled thickness. Despite the simplicity and significance of creating a new class of materials, namely 2D covalent solids, this concept has not yet been explored for atomically thin layers of van der Waals crystals. Here, we successfully demonstrate the validity of this approach by reporting the chemical conversion of three or more layers of InSe into covalent $InF_3$ thin films by fluorination of layered 2D crystals of InSe.

We used the mechanical exfoliation method to prepare InSe flakes on a quartz substrate[6, 18, 19]. Exfoliated InSe flakes and bulk InSe crystals were then fluorinated by direct exposure to Xenon difluoride ($XeF_2$) at elevated temperature using a method reported earlier[6] (see methods). Figure 1a shows the optical micrographs of InSe flakes before and after the fluorination process. After fluorination, the optical contrast from the thinner InSe flakes decreases significantly, but atomic force microscopy (AFM) confirms the preservation of the flake morphology and its 2D structure (Figure 1b). Using this method, we have fluorinated InSe flakes down to three-layer thicknesses (2.4 nm) and obtained fluorinated flakes with a reduced thickness of 1.5 nm (Figure S1). The fluorination of mono- and bi-layer InSe resulted in the disappearance of the flakes, indicating an inherent instability of the thinner crystals. It is also noteworthy that thicker layers of InSe preserved their anisotropic structure after fluorination (Figure 1a,b).

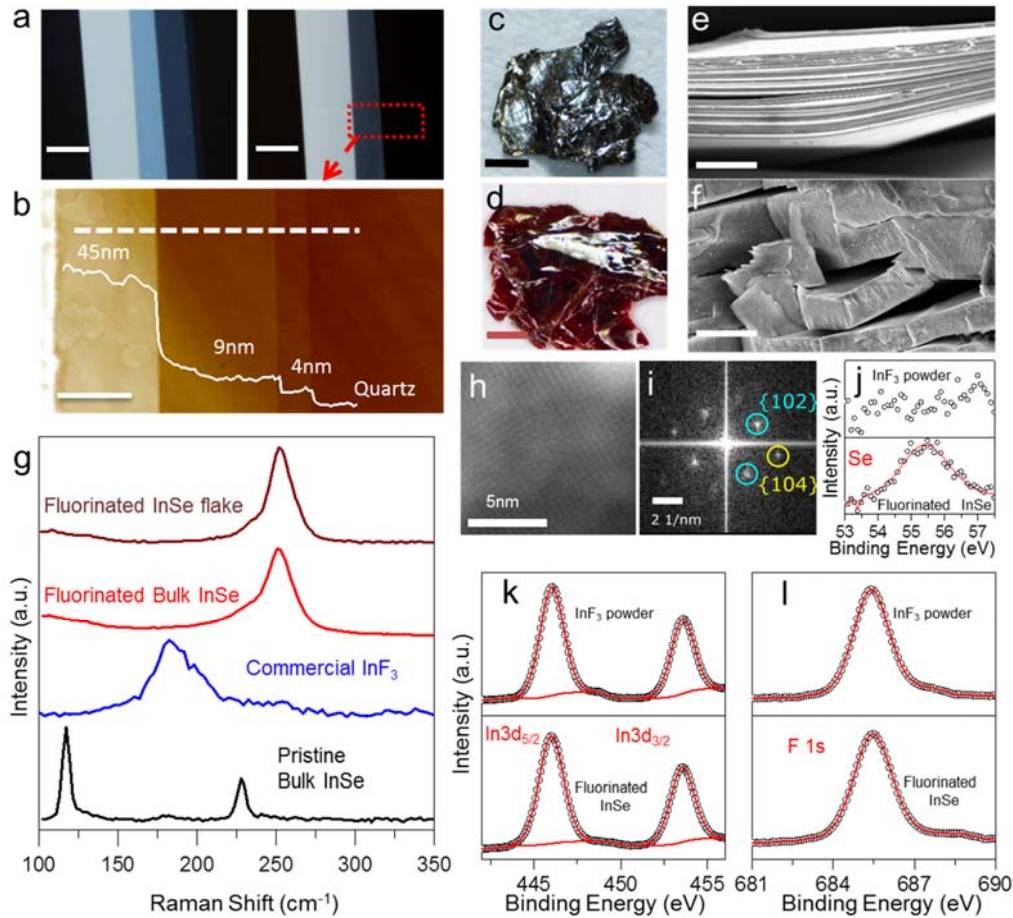

**Figure 1. Characterization of fluorinated InSe. a**, Optical microscope images of InSe flakes on quartz substrate before (top left) and after (top right) fluorination. Scale bars, 7 μm. **b**, An AFM image of the area marked with the red rectangle in Figure 1a. Scale bar, 5 μm. White curve: height profile along the dashed line. **c, and d,** Photographs of bulk InSe before and after the fluorination, respectively. Scale bars, 1 mm. **e, and f,** Cross-sectional SEM images of bulk pristine InSe and fluorinated InSe, respectively. Scale bars, 5 μm. **g**, Raman spectra of a fluorinated InSe flake (≈10 nm thick), fluorinated bulk InSe, commercial $InF_3$, and pristine bulk InSe. **h**, High angle annular dark field (HAADF) scanning transmission electron microscopy (STEM) image of fluorinated InSe. Scale bar, 5 nm. **i**, Fast Fourier transform (FFT) from the region in Figure 1h, showing {102}

and {104} planes in k-space. Scale bar, 2 nm$^{-1}$. XPS spectra of bulk fluorinated InSe crystal and commercial InF$_3$ powder showing **j**, selenium, **k**, indium, and **l**, fluorine peaks.

Figure 1c,d shows the optical photographs of pristine and fluorinated bulk InSe crystals. The fluorinated bulk crystals are red in colour (Figure 1d), in sharp contrast to bulk InSe which is black (Figure 1c). After fluorination, the crystal becomes harder and brittle compared to the pristine InSe, indicating the loss of layered structure and formation of a covalent crystal. Direct exfoliation of these fluorinated crystals using micromechanical exfoliation was not successful, confirming the covalent nature of the InF$_3$ crystal. This is further confirmed by the cross-sectional scanning electron microscopy characterisation shown in Figure 1e,f where a layered to a non-layered transition is apparent after fluorination.

We used Raman spectroscopy, transmission electron microscopy (TEM), and X-ray photoelectron spectroscopy (XPS) to characterize the fluorinated samples. The Raman spectrum of bulk InSe shows three characteristic peaks at 117 cm$^{-1}$, 179 cm$^{-1}$ and 227 cm$^{-1}$ (Figure 1g) corresponding to the A$_1'$($\Gamma_1^2$), E$'$($\Gamma_3^1$) – TO + E$''$($\Gamma_3^3$) and A$_1'$($\Gamma_1^3$) phonon modes[20]. These peaks are not observed after fluorination. Instead, the fluorinated samples show a peak at ≈ 250 cm$^{-1}$ with a shoulder at 232 cm$^{-1}$ and another weaker mode at ≈495 cm$^{-1}$ (Figure 1g and Figure S2). The absence of the InSe Raman peaks and the appearance of new peaks in the Raman spectra suggest complete fluorination of both the bulk InSe and the InSe flakes on quartz. Moreover, similarities between the Raman spectra from the bulk fluorinated crystal and the exfoliated flake suggest similar fluorination chemistry for both types of samples. Due to the relatively high affinity of indium to fluorine, one would expect the most probable outcome of the InSe fluorination to be InF$_3$[21]. High-resolution transmission electron microscopy (HRTEM) of the fluorinated InSe samples is consistent with the InF$_3$ atomic structure (Figure 1h,i), with the Fourier transform of the HRTEM image showing the {102} and {104} planes of the InF$_3$ crystal (R-3c h (167) space group) with lattice spacings of 3.93 Å and 2.854 Å, respectively[22]. In agreement with this, the selected area electron diffraction and the X-ray diffraction experiments (Figure S3 and S4) further confirm the expected InF$_3$ structure for fluorinated InSe.

It is notable, however, that the Raman spectra for commercial InF$_3$ powders were found to be different from those of the fluorinated InSe crystals, indicating that these materials might have a different chemical composition (Figure 1g). The difference in the Raman spectrum of the reference

InF$_3$ and the InF$_3$ obtained by fluorination of InSe is intriguing. To understand this difference, we conducted an XPS analysis of these samples. Figure 1j-l shows XPS from the reference commercial InF$_3$ and the InF$_3$ obtained by fluorination of InSe. Both samples show indium 3d$_{3/2}$, and 3d$_{5/2}$ peaks at 453.5 eV and 446 eV, along with the fluorine 1s peak at 685.5 eV, which confirms the covalent InF$_3$ atomic structure of the fluorinated samples, in agreement with electron microscopy and X-ray analysis[6]. Furthermore, a closer look of the spectra from the fluorinated InSe reveals a significant amount (≈2%, see methods) of selenium doping in the crystal (Figure 1j). This doping was absent in the commercial InF$_3$ samples (Figure 1j) and could be responsible for the variation in the Raman spectra compared to commercial InF$_3$ powder, as discussed below. We have also noticed that the Raman spectrum of Se-doped InF$_3$ was not affected by slight changes in the amount of Se doping (Figure S5).

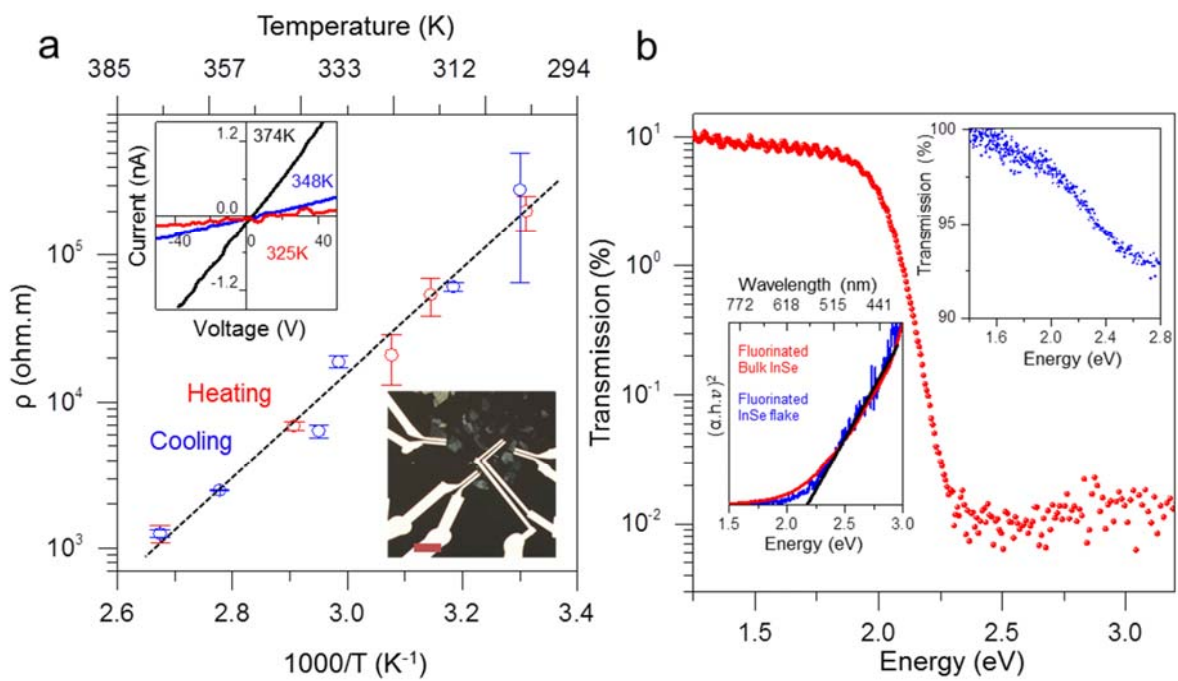

**Figure 2. Band gap estimation. a**, Temperature-dependent electrical resistivity of the fluorinated InSe flake. Top inset: Current-voltage I-V curves at different temperatures (colour coded labels). Bottom inset: optical micrograph of the device. Scale bar, 10 μm. **b**, Optical transmission through fluorinated bulk InSe and fluorinated InSe flake (≈10 nm in thickness) on a quartz substrate (top inset) at 300K. Bottom inset: the associated Tauc plots (see methods) indicate a direct bandgap of 2.2 eV for both bulk fluorinated InSe and the fluorinated InSe flake (colour coded labels).

To assess the electronic properties of the Se-doped InF$_3$ films, we performed electrical and optical measurements (see methods). Electrical transport measurements were performed on two-terminal devices fabricated with a fluorinated InSe flake (≈10 nm thick) on a quartz substrate (see methods). These devices showed a high room-temperature resistivity, typically in the range of 10$^5$ Ω.m (Figure 2a). The temperature dependent electrical resistivity follows an Arrhenius-type behaviour described by the empirical relation

$$\rho = \rho_0 \cdot e^{\left(\frac{E_A}{k_B T}\right)} \qquad (1)$$

where $E_A$ is an activation energy, $k_B$ is the Boltzmann constant, $T$ is the temperature in Kelvin, and the parameter $\rho_0$ is only weakly dependent on $T$ (Figure 2a)[6, 23]. The fitting of the data (shown in Figure 2a) yields $E_A = 0.7 \pm 0.1$ eV. The electrical measurements revealed no change in the room temperature resistivity after heating the fluorinated samples in air to temperatures of 100 °C and subsequent cooling down, confirming the thermal stability of the Se-doped InF$_3$.

For an intrinsic semiconductor, the activation energy $E_A$ from the transport measurements would be related to the bandgap energy $E_g$ of the material, i.e. $E_g/2 = E_A$. On the other hand, for an extrinsic semiconductor containing impurities and/or defects in the lattice, $E_A$ would be related to the binding energy and occupancy of the defect/impurity states within the bandgap[6]. To probe the bandgap energy of the fluorinated samples (Se doped InF$_3$), we performed optical absorption experiments (see methods). Figure 2b shows the optical transmission spectra of the Se-doped InF$_3$ crystals, demonstrating a strong optical absorption above ≈2 eV. The Tauc plot analysis (Figure 2b bottom inset; see methods) reveals that the Se-doped InF$_3$ samples obtained by fluorination of both InSe flakes and bulk crystals have a direct bandgap with $E_g ≈ 2.2$ eV [3]. This value is significantly larger than the energy $2E_A = 1.4$ eV derived from the electrical measurements, suggesting that the chemical potential is pinned on mid-gap impurity states. We have also investigated the photoluminescence of the Se-doped InF$_3$, but no emission was observed, consistent with the presence of non-radiative recombination centres, as was reported earlier for fluorographene[6]. We also calculated the optical constants of the Se-doped InF$_3$, which showed an almost constant real refractive index of 2, while the imaginary part of the refractive index reduced to near zero for wavelengths > 600 nm (Figure S6). This implies that Se-doped InF$_3$ transmits all wavelengths above 600 nm with near zero attenuation.

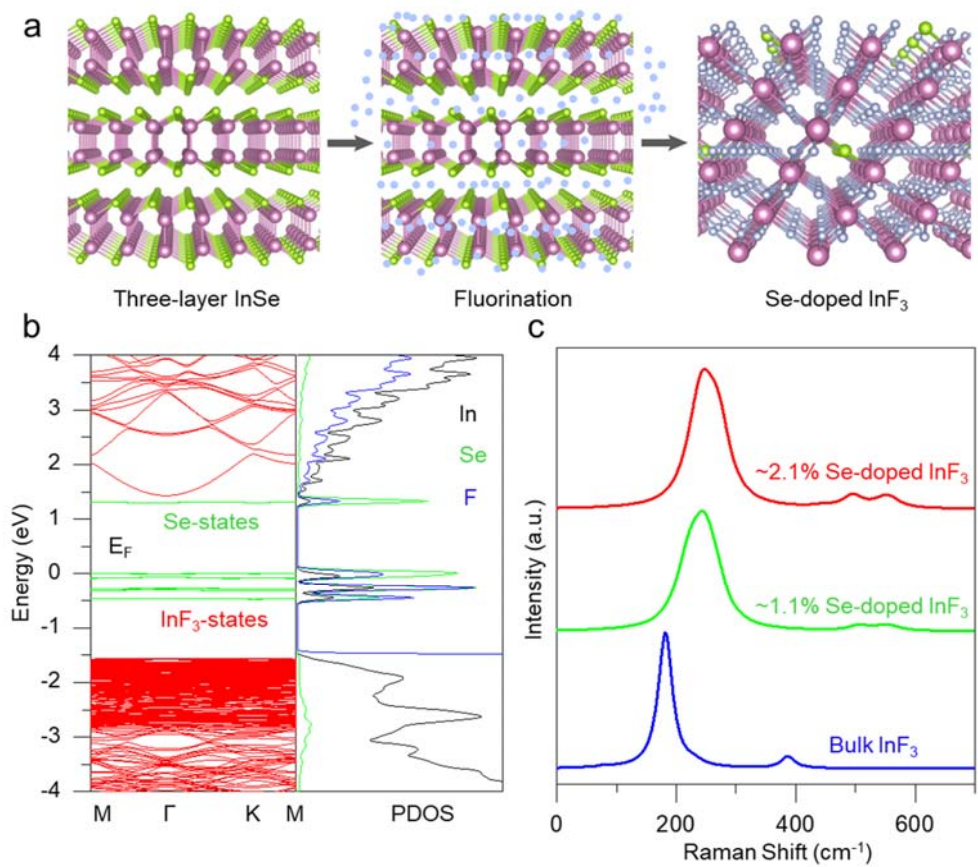

**Figure 3. *Ab-initio* density functional theory calculations. a**, Schematic showing the fluorination of three-layer InSe to $InF_3$. Purple, light blue, and green spheres corresponds to indium, fluorine, and selenium, respectively. **b**, Electronic-band structure and the corresponding partial density of states (PDOS) for 2.1% Se-doped $InF_3$ (color-coded labels). The green bands represent Se states. The midgap fluorine and indium states are absent in the pristine $InF_3$ (Figure S8) and only introduced after Se doping. The Fermi energy is set to 0 eV. **c**, Calculated Raman spectrum of bulk $InF_3$ and Se-doped $InF_3$ at two different doping levels.

To further understand the fluorination of InSe, we performed *ab-initio* Density Functional Theory (DFT) calculations (see methods). Our structure optimization and phonon dispersion calculations show that a minimum of three layers of InSe is essential for the formation of $InF_3$ crystal by fluorination of InSe (Figure S7). Fewer than three layers of fully fluorinated InSe were dynamically unstable and did not structurally converge into $InF_3$, in agreement with the experimental findings. Figure 3a shows a schematic diagram of the chemical conversion of 3 layers of InSe into Se-doped $InF_3$. Electronic band dispersion calculations reveal that bulk $InF_3$ is a wide gap semiconductor with a bandgap of 2.9 eV (Figure S8). The doping of $InF_3$ with Se results in impurity mid-gap states, which significantly decreases the band gap of InF$_3$ to ≈2.2 eV, as observed in the experiment. As shown by the green bands in Figure 3b, the impurity mid-gap states arise almost entirely from the Se orbital states, giving rise to a smaller bandgap in Se-doped InF$_3$. Also, as shown in Figure 3b, the band dispersions (red bands) of both the valence band and the conduction band edges of InF$_3$ are mostly dominated by the F-*p* orbitals, with the In-*d* orbitals also contributing to the valence band edges.

We also modelled the vibrational spectrum of InF$_3$ and Se-doped InF$_3$, and found that Se-doping modifies the Raman spectrum of InF$_3$, consistent with the experimental results. The calculated Raman spectrum of both InF$_3$ and Se-doped InF$_3$ (Figure 3c) agrees well with the experimentally recorded spectrum (see supplementary text 1). We have also noticed that the calculated Raman spectrum of Se-doped InF$_3$ was not affected by slight changes in the amount of Se doping as seen in experiment (Fig. 3c and Figure S5).

To demonstrate the scalability of our synthesis procedure, we prepared large area Se-doped InF$_3$ by fluorinating liquid-exfoliated InSe laminates (see methods and Figure S9 and S10). Liquid exfoliated InSe flakes (≈2.5 nm thick) were obtained by ultra-sonication of InSe powder in Isopropyl Alcohol[24,25]. These nanosheets were filtered onto an Anodisc alumina/PTFE membrane and fluorinated at 350 °C for 48 hrs. The resulting InF$_3$ films show structural and optical characteristics similar to Se-doped InF$_3$ obtained by the fluorination of mechanically exfoliated InSe (Figure S10).

In conclusion, we have successfully demonstrated the synthesis of atomically thin covalent crystals by chemical conversion of a few-layered van der Waals material. This is achieved by synthesizing atomically thin Se-doped InF$_3$ by fluorinating few-layer InSe flakes. The resulting Se-doped InF$_3$ was found to be a semiconductor with an optical band gap of ≈2.2 eV. The experimentally observed Raman spectra and bandgap match the calculated values. Our approach could be used to synthesize a large variety of other 2D covalent solids, which cannot be produced by exfoliation.

**Methods:**

**Fluorination of InSe flakes:** InSe flakes were prepared on a quartz wafer using micromechanical exfoliation of Bridgman-grown bulk γ-InSe crystals. Quartz substrates of size (≈5 cm x 5 cm) were cleaned using acetone and isopropyl alcohol for 10 min each, followed by $O_2$/Ar plasma cleaning for 10 min. The substrates were immediately transferred inside a glove box for mechanical exfoliation of InSe crystals under argon environment with $H_2O$ and $O_2$ levels less than 0.1 ppm. The transferred crystals on quartz were then taken to a PTFE container mixed with excess $XeF_2$ crystals and heated to 100 °C on a hot plate for 48 hrs. After fluorination the sample was then annealed at 80 °C for 12 hrs inside a glove box to remove residual $XeF_2$ crystals and used for subsequent measurements.

**Fluorination of bulk InSe:** A small piece of InSe crystal was scrapped off a large crystal using a surgical blade inside a glove box. The crystal was then mixed with excess (~ 100 times by mass) $XeF_2$ inside a PTFE lined stainless steel container and heated to 350 °C for 48 hrs on a hot plate. The crystals thus obtained were annealed at 150 °C for 12 hrs inside a glove box to remove the residual $XeF_2$ crystals and used for subsequent measurements.

**Preparation and fluorination of InSe laminates:** InSe powder for liquid exfoliation was purchased from Chengdu Alfa Metal Material Co.,Ltd, China and used as received. 10 mg of InSe powder was dispersed in about 100 ml of iso-propyl alcohol. The dispersion was then sonicated for 12 hrs. The resulting dispersion after sonication was centrifuged at 4000 rpm for 10 min. The supernatant was collected and filtered through a PTFE (0.1 μm pore size and 47 mm diameter, purchased from Sterlitech) or Anodisc alumina membrane (0.2 μm pore size and a diameter of 47 mm, purchased from Millipore) to obtain InSe laminates with thickness of a few microns. These laminates were fluorinated at 350 °C on a hot plate in a PTFE-lined autoclave for 48 hrs and characterised using X-ray diffraction. After the fluorination, the $InF_3$ coating was loosely attached to the PTFE/alumina substrate and could be peeled from the substrate to obtain freestanding films of $InF_3$ (Figure S10).

**AFM measurements:** AFM imaging was performed using a Bruker Dimension FastScan AFM operating in peak force tapping mode.

**Raman measurements:** We used HORIBA's Raman spectrometer (XploRA PLUS) with a laser excitation of 532 nm (spot size ~1 μm, laser power of 1.35 mW and spectrometer grating of 1200 groves per millimetre) for measuring the Raman spectra of the samples.

**TEM measurements:** Samples for TEM were prepared by rubbing $InF_3$ bulk crystals against a TEM grid (metal mesh). We used an FEI Titan G2 80-200 S/TEM 'ChemiSTEM' operated at 200 kV. Imaging was carried out in HAADF STEM mode, with a probe current of 180 pA and convergence angle 21 mrad, while diffraction was performed in TEM mode.

**X-Ray Diffraction measurements:** X-ray diffraction (XRD) was performed using a Rigaku SmartLab XRD with Cu Kα radiation (medium resolution parallel beam measurement mode, λ = 1.5406 Å) with a step size of 0.01 degrees. The crystals were pressed and made into a powder using a mortar before placing into the X-ray sample holder. The voltage and current of the X-ray tube was fixed to 40 KV and 45 mA, respectively. XRD from liquid exfoliated InSe before and after fluorination was obtained directly from the as-prepared samples.

**XPS measurements:** XPS analysis was performed with a Kratos AXIS Ultra DLD apparatus, equipped with a monochromated Al Kα radiation X-ray source, a charge neutralizer, and a hemispherical electron energy analyser. During data acquisition, the chamber pressure was kept below $10^{-9}$ mbar. The spectra were analysed using the CasaXPS software pack and corrected for charging using C 1s binding energy as the reference at 284.8 eV. Survey scans and high-resolution scans were carried out at pass energies of 80 eV and 20 eV, respectively. Atomic percentage of elements were calculated from the survey scans using CasaXPS software pack.

**Electrical measurements:** Electrical measurements of $InF_3$ crystals were conducted using a 2400 Keithley source meter. The temperature-dependent conductivity studies were conducted by heating a device in air using a hot plate. The temperature on the surface of the sample was measured using an infrared thermometer.

**Optical measurements:** Wavelength-dependent transmission of $InF_3$ samples was measured using a homemade spectrometer with focusing optics in transmission mode. The incident light from a laser driver light source (LDLS) was focused on the sample with the FL 40x objective and then collected using a similar objective. The transmitted light was focused on the entrance of an optical fiber (200 μm core) coupled to the Ocean Optics USB2000 spectrometer. The transmission spectra were obtained by normalizing the spectra measured from the sample with respect to the spectra measured from the substrate. We extracted the spectral dependencies of the complex refractive index using spectroscopic ellipsometry. The ellipsometric measurements were performed with a Woollam VASE variable angle ellipsometer (M-2000F) with a focal spot of just 30 μm in the wavelength range of 240–1700 nm. The ellipsometric data were modelled with WVASE32 software based on Fresnel coefficients for multilayered films.

**Bandgap calculations:** To probe the optical band gap of the Se-doped InF$_3$ crystals, we used the measured absorption coefficient (Figure S6) to construct Tauc plots. The optical absorption strength depends on the difference between the photon energy and the bandgap as

$$(\alpha h v)^{1/n} = A(hv - E_g) \qquad (2)$$

where $h$ is Planck's constant, $v$ is the photon's frequency, $\alpha$ is the absorption coefficient, $A$ is proportionality constant, and $E_g$ is the bandgap. $E_g$ is obtained from the X-intercept of the linear regime of the Tauc plot $[(\alpha hv)^{\frac{1}{n}}$ vs. $hv]$. The nature of the transition is denoted by the value of $n$, where $n = \frac{1}{2}, \frac{3}{2}, 2$ or $3$ for direct allowed, direct forbidden, indirect allowed, and indirect forbidden transitions, respectively. For our samples, n = $\frac{1}{2}$ provided the best linear regime in the Tauc plot (Figure 2b inset).

**DFT-methodology:** For the investigation of the structural, electronic, and vibrational properties of bulk InSe and doped InF$_3$ crystals, density functional theory (DFT)-based first principle calculations were performed as implemented in the Vienna ab-initio simulation package (VASP)[26]. The Perdew-Burke-Ernzerhof (PBE) form of generalized gradient approximation (GGA) was adopted to describe electron exchange and correlation[27]. In order to take into account strong correlations between In-$d$ orbital electrons, The DFT+U method was used[28]. The effective on-site Coulomb parameter, U$_{eff}$, was taken to be 7 eV.

The kinetic energy cutoff for plane-wave expansion was set to 500 eV and the energy was minimized until its variation in the following steps became less than 10$^{-8}$ eV. The width of the Gaussian smearing was chosen to be 0.05 eV for both geometry optimizations and partial density of states (PDOS) calculations. For the vibrational spectrum of each bulk structure, first-order off-resonant Raman activities were calculated at the Γ point of the Brillouin Zone (BZ). Firstly, the zone-centered phonon modes were calculated using small-displacement methodology as implemented in VASP. Then using the vibrational characteristic of each optical phonon mode, the derivative of macroscopic dielectric tensor was calculated with respect to each normal vibrational mode to obtain the corresponding Raman activity[29, 30].

**ACKNOWLEDGMENT:** This work was supported by the Royal Society, the European Research Council (contract 679689 and EvoluTEM 715502), and Engineering and Physical Sciences

Research Council, UK (EP/N013670/1). The authors acknowledge the use of the facilities at the Henry Royce Institute for Advanced Materials and associated support services. H.S. acknowledges financial support from the Scientific and Technological Research Council of Turkey (TUBITAK) under Project No. 117F095. M.Y. acknowledges the Flemish Science Foundation (FWO-Vl) for a postdoctoral fellowship. SJH and DJK acknowledge support from EPSRC (EP/P009050/1) and the NowNANO CDT.

**REFERENCES**

(1) Lei, S.; Wang, X.; Li, B.; Kang, J.; He, Y.; George, A.; Ge, L.; Gong, Y.; Dong, P.; Jin, Z.; Brunetto, G.; Chen, W.; Lin, Z.-T.; Baines, R.; Galvão, D. S.; Lou, J.; Barrera, E.; Banerjee, K.; Vajtai, R.; Ajayan, P., Surface functionalization of two-dimensional metal chalcogenides by Lewis acid–base chemistry. *Nat. Nanotech.* **2016**, 11, 465.

(2) Radhakrishnan, S.; Das, D.; Deng, L.; Sudeep, P. M.; Colas, G.; de los Reyes, C. A.; Yazdi, S.; Chu, C. W.; Martí, A. A.; Tiwary, C. S.; Filleter, T.; Singh, A. K.; Ajayan, P. M., An Insight into the Phase Transformation of $WS_2$ upon Fluorination. *Adv. Mater.* **2018**, 30, 1803366.

(3) Radhakrishnan, S.; Das, D.; Samanta, A.; de los Reyes, C. A.; Deng, L.; Alemany, L. B.; Weldeghiorghis, T. K.; Khabashesku, V. N.; Kochat, V.; Jin, Z.; Sudeep, P. M.; Martí, A. A.; Chu, C.-W.; Roy, A.; Tiwary, C. S.; Singh, A. K.; Ajayan, P. M., Fluorinated h-BN as a magnetic semiconductor. *Sci. Adv.* **2017**, 3, e1700842.

(4) Voiry, D.; Goswami, A.; Kappera, R.; Silva, C. d. C. C. e.; Kaplan, D.; Fujita, T.; Chen, M.; Asefa, T.; Chhowalla, M., Covalent functionalization of monolayered transition metal dichalcogenides by phase engineering. *Nat. Chem.* **2014**, 7, 45.

(5) Elias, D. C.; Nair, R. R.; Mohiuddin, T. M. G.; Morozov, S. V.; Blake, P.; Halsall, M. P.; Ferrari, A. C.; Boukhvalov, D. W.; Katsnelson, M. I.; Geim, A. K.; Novoselov, K. S., Control of Graphene's Properties by Reversible Hydrogenation: Evidence for Graphane. *Science* **2009**, 323, 610-613.

(6) Nair, R. R.; Ren, W.; Jalil, R.; Riaz, I.; Kravets, V. G.; Britnell, L.; Blake, P.; Schedin, F.; Mayorov, A. S.; Yuan, S.; Katsnelson, M. I.; Cheng, H.-M.; Strupinski, W.; Bulusheva, L. G.; Okotrub, A. V.; Grigorieva, I. V.; Grigorenko, A. N.; Novoselov, K. S.; Geim, A. K., Fluorographene: A Two-Dimensional Counterpart of Teflon. *Small* **2010**, 6, 2877-2884.

(7) Geim, A. K.; Grigorieva, I. V., Van der Waals heterostructures. *Nature* **2013**, 499, 419.

(8) Carvalho, A.; Wang, M.; Zhu, X.; Rodin, A. S.; Su, H.; Castro Neto, A. H., Phosphorene: from theory to applications. *Nat. Rev. Mater.* **2016**, 1, 16061.

(9) Bhimanapati, G. R.; Lin, Z.; Meunier, V.; Jung, Y.; Cha, J.; Das, S.; Xiao, D.; Son, Y.; Strano, M. S.; Cooper, V. R.; Liang, L.; Louie, S. G.; Ringe, E.; Zhou, W.; Kim, S. S.; Naik, R. R.; Sumpter, B. G.; Terrones, H.; Xia, F.; Wang, Y.; Zhu, J.; Akinwande, D.; Alem, N.; Schuller, J. A.; Schaak, R. E.; Terrones, M.; Robinson, J. A., Recent Advances in Two-Dimensional Materials beyond Graphene. *ACS Nano* **2015**, 9, 11509-11539.

(10)    Puthirath Balan, A.; Radhakrishnan, S.; Woellner, C. F.; Sinha, S. K.; Deng, L.; Reyes, C. d. l.; Rao, B. M.; Paulose, M.; Neupane, R.; Apte, A.; Kochat, V.; Vajtai, R.; Harutyunyan, A. R.; Chu, C.-W.; Costin, G.; Galvao, D. S.; Martí, A. A.; van Aken, P. A.; Varghese, O. K.; Tiwary, C. S.; Malie Madom Ramaswamy Iyer, A.; Ajayan, P. M., Exfoliation of a non-van der Waals material from iron ore hematite. *Nat. Nanotech.* **2018,** 13, 602-609.

(11)    Dávila, M. E.; Xian, L.; Cahangirov, S.; Rubio, A.; Le Lay, G., Germanene: a novel two-dimensional germanium allotrope akin to graphene and silicene. *New J. Phys.* **2014**, 16, 095002.

(12)    Tao, L.; Cinquanta, E.; Chiappe, D.; Grazianetti, C.; Fanciulli, M.; Dubey, M.; Molle, A.; Akinwande, D., Silicene field-effect transistors operating at room temperature. *Nat. Nanotech.* **2015**, 10, 227.

(13)    Zhu, F.-f.; Chen, W.-j.; Xu, Y.; Gao, C.-l.; Guan, D.-d.; Liu, C.-h.; Qian, D.; Zhang, S.-C.; Jia, J.-f., Epitaxial growth of two-dimensional stanene. *Nat. Mater.* **2015**, 14, 1020.

(14)    Mannix, A. J.; Zhang, Z.; Guisinger, N. P.; Yakobson, B. I.; Hersam, M. C., Borophene as a prototype for synthetic 2D materials development. *Nat. Nanotech.* **2018**, 13, 444-450.

(15)    Chen, H.; Chen, Z.; Ge, B.; Chi, Z.; Chen, H.; Wu, H.; Cao, C.; Duan, X., General Strategy for Two-Dimensional Transition Metal Dichalcogenides by Ion Exchange. *Chem. Mater.* **2017**, 29, 10019-10026.

(16)    Bouet, C.; Laufer, D.; Mahler, B.; Nadal, B.; Heuclin, H.; Pedetti, S.; Patriarche, G.; Dubertret, B., Synthesis of Zinc and Lead Chalcogenide Core and Core/Shell Nanoplatelets Using Sequential Cation Exchange Reactions. *Chem. Mater.* **2014**, 26, 3002-3008.

(17)    Wang, Y.; Morozov, Y. V.; Zhukovskyi, M.; Chatterjee, R.; Draguta, S.; Tongying, P.; Bryant, B.; Rouvimov, S.; Kuno, M., Transforming Layered to Nonlayered Two-Dimensional Materials: Cation Exchange of $SnS_2$ to $Cu_2SnS_3$. *ACS Energy Lett.* **2016**, 1, 175-181.

(18)    Bandurin, D. A.; Tyurnina, A. V.; Yu, G. L.; Mishchenko, A.; Zólyomi, V.; Morozov, S. V.; Kumar, R. K.; Gorbachev, R. V.; Kudrynskyi, Z. R.; Pezzini, S.; Kovalyuk, Z. D.; Zeitler, U.; Novoselov, K. S.; Patanè, A.; Eaves, L.; Grigorieva, I. V.; Fal'ko, V. I.; Geim, A. K.; Cao, Y., High electron mobility, quantum Hall effect and anomalous optical response in atomically thin InSe. *Nat. Nanotech.* **2016**, 12, 223.

(19)    Mudd, G. W.; Svatek, S. A.; Ren, T.; Patanè, A.; Makarovsky, O.; Eaves, L.; Beton, P. H.; Kovalyuk, Z. D.; Lashkarev, G. V.; Kudrynskyi, Z. R.; Dmitriev, A. I., Tuning the Bandgap of Exfoliated InSe Nanosheets by Quantum Confinement. *Adv. Mater.* **2013**, 25, 5714-5718.

(20)    Lei, S.; Ge, L.; Najmaei, S.; George, A.; Kappera, R.; Lou, J.; Chhowalla, M.; Yamaguchi, H.; Gupta, G.; Vajtai, R.; Mohite, A. D.; Ajayan, P. M., Evolution of the Electronic Band Structure and Efficient Photo-Detection in Atomic Layers of InSe. *ACS Nano* **2014**, 8, 1263-1272.

(21)    Davidovich Ruven, L.; Fedorov Pavel, P.; Popov Artur, I., Structural chemistry of anionic fluoride and mixed-ligand fluoride complexes of indium(III). *Rev. Inorg. Chem.* **2016,** 36, 105.

(22)    P. Villars, PAULING FILE in: Inorganic Solid Phases, SpringerMaterials (online database) *(http://materials.springer.com/isp/crystallographic/docs/sd_0541715)*, Springer, Heidelberg (ed.) Springer Materials, 2012 (Date accessed May, 10, 2019).

(23)    Roy, T. K.; Sanyal, D.; Bhowmick, D.; Chakrabarti, A., Temperature dependent resistivity study on zinc oxide and the role of defects. *Mater. Sci. Semicond. Process.* **2013**, 16, 332-336.

(24) Nicolosi, V.; Chhowalla, M.; Kanatzidis, M. G.; Strano, M. S.; Coleman, J. N., Liquid Exfoliation of Layered Materials. *Science* **2013**, 340, 1226419.

(25) Petroni, E.; Lago, E.; Bellani, S.; Boukhvalov, D. W.; Politano, A.; Gürbulak, B.; Duman, S.; Prato, M.; Gentiluomo, S.; Oropesa-Nuñez, R.; Panda, J.-K.; Toth, P. S.; Del Rio Castillo, A. E.; Pellegrini, V.; Bonaccorso, F., Liquid-Phase Exfoliated Indium–Selenide Flakes and Their Application in Hydrogen Evolution Reaction. *Small* **2018**, 14, 1800749.

(26) Kresse, G.; Furthmüller, J., Efficient iterative schemes for ab initio total-energy calculations using a plane-wave basis set. *Phys. Rev. B* **1996**, 54, 11169-11186.

(27) Perdew, J. P.; Burke, K.; Ernzerhof, M., Generalized Gradient Approximation Made Simple. *Phys. Rev. Lett.* **1996**, 77, 3865-3868.

(28) Dudarev, S. L.; Botton, G. A.; Savrasov, S. Y.; Humphreys, C. J.; Sutton, A. P., Electron-energy-loss spectra and the structural stability of nickel oxide: An LSDA+U study. *Phys. Rev. B* **1998**, 57, 1505-1509.

(29) Yagmurcukardes, M.; Peeters, F. M.; Sahin, H., Electronic and vibrational properties of PbI2: From bulk to monolayer. *Phys. Rev. B* ***2018***, 98, 085431.

(30) Yagmurcukardes, M.; Bacaksiz, C.; Unsal, E.; Akbali, B.; Senger, R. T.; Sahin, H., Strain mapping in single-layer two-dimensional crystals via Raman activity. *Phys. Rev. B* **2018**, 97, 115427.

# Supplementary Information

**Supplementary text**

**#1. Calculated Raman Spectrum of InF$_3$ and Se doped InF$_3$**

The calculated Raman spectrum of InF$_3$ reveals that it has a prominent peak located at 182.1 cm$^{-1}$, consistent with the experimental data, which is attributed to the in-plane vibrations of the F-atoms. Apart from this most prominent peak, there are some low intensity Raman-active modes at 230.7, 249.1, 251.2, 252.5, 259.6, 386.6, and 474.4 cm$^{-1}$ (Figure. 3c). Our calculation reveals that Se-doping results in additional Raman active modes. For a Se doping of 2.1%, a broad prominent peak centred at 250.6 cm$^{-1}$ is observed. Note that in the pure InF$_3$ structure, the Raman modes with frequencies 249.1, 251.2, 252.5, and 259.6 cm$^{-1}$ are non-degenerate, whereas in the Se-doped InF$_3$ structure, due to the 2×2×1 supercell, each mode has 4-fold-degeneracy. However, due to the small distortion in the supercell, the degeneracy of these modes is broken and 16 phonon modes appear, some of which are Raman inactive. Therefore, in the Raman spectrum of the Se-doped structure, the phonon modes around the broad peak at 250 cm$^{-1}$ are attributed to the modes arising from the InF$_3$ crystal. Moreover, the shoulder peak at 228.6 cm$^{-1}$ (experimentally found to be around ≈232 cm$^{-1}$) is attributed to the In-Se bond stretching. We have also noticed weaker modes around ≈495 cm$^{-1}$ that arise from bulk InF$_3$. Their intensities are very low and only become prominent after Se-doping (Figure. S2).

**Supplementary Figures**

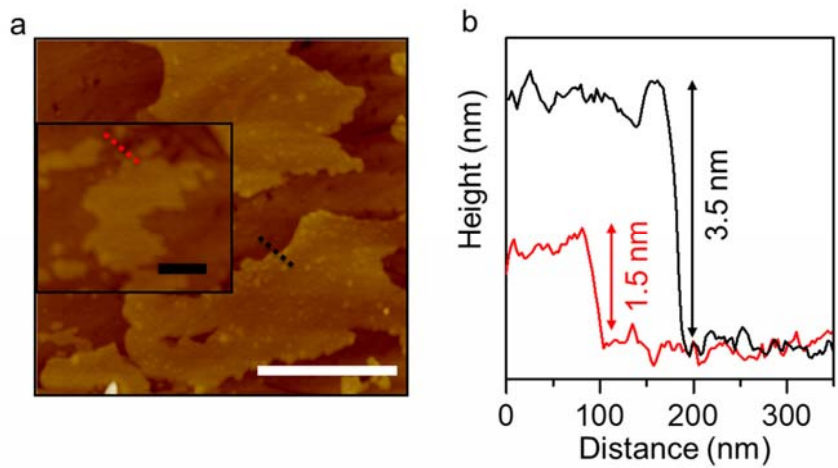

**Figure S1. Atomic Force Microscopy (AFM) Image of thin InF$_3$ crystals.** **a**, AFM image of 4 nm thick and 1.5 nm thick (inset. Scale bar, 0.5 μm) fluorinated InSe flakes. Scale bar, 1.5 μm. **b**, The height profile along the dotted lines in Figure S1a.

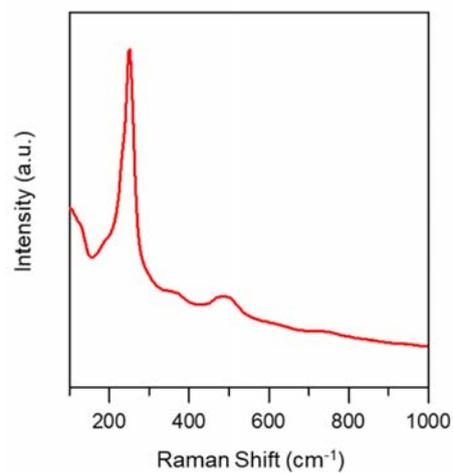

**Figure S2. Raman spectrum of fluorinated InSe.** Raman spectrum of fluorinated InSe over an extended range of wavenumbers.

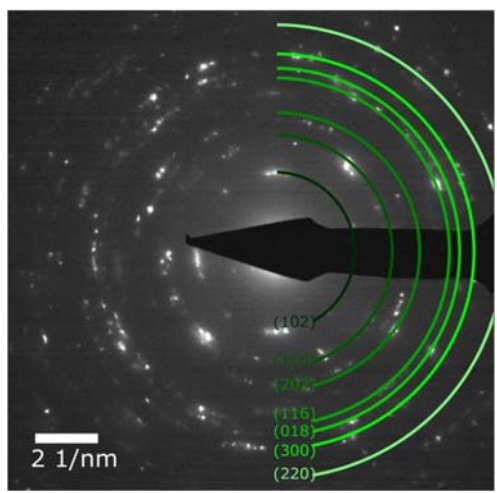

**Figure S3. Selected Area Electron Diffraction (SAED) pattern of fluorinated InSe.** SAED pattern obtained from a polycrystalline region of the fluorinated InSe sample. The labelled diffraction rings correspond to specific planes of the $InF_3$ crystal structure.

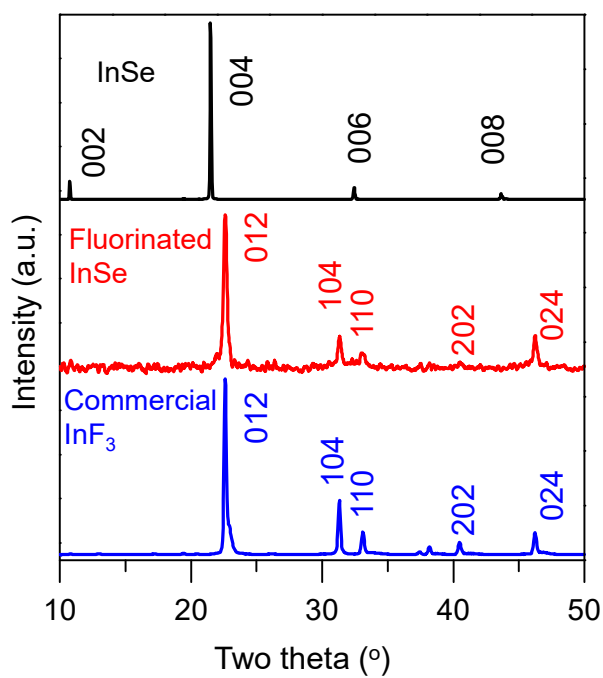

**Figure S4. X-Ray Diffraction (XRD) characterization of fluorinated InSe.** XRD spectra of pristine InSe, InSe after fluorination, and commercial $InF_3$ powder (colour-coded labels). The similar XRD spectra of fluorinated InSe and $InF_3$ indicate the conversion of InSe into $InF_3$ after the fluorination. The absence of any InSe diffraction peak in the fluorinated sample confirms the complete conversion of InSe into $InF_3$. The Miller indices of the Bragg reflections are labelled.

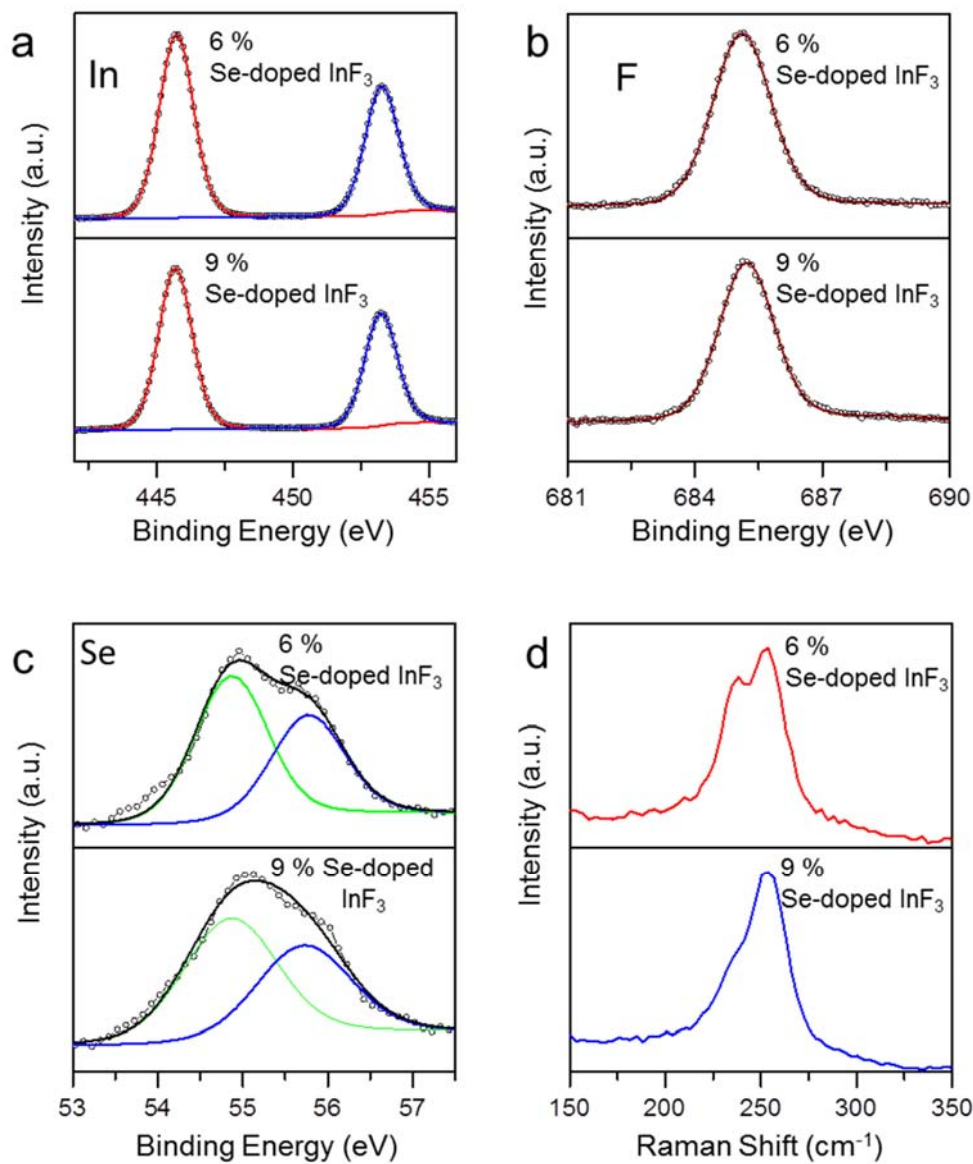

**Figure S5. Tuning Se doping in InF$_3$ samples**. XPS spectra of 6% and 9% Se doped InF$_3$ showing **a,** Indium, **b,** Fluorine and **c,** Selenium peaks. Se doping levels in the samples were controlled by the amount of XeF$_2$ crystals used for fluorination; higher XeF$_2$ leads to smaller doping level (~10 times by mass gives 9% Se doping and 40 times by mass gives 6% Se doping) .**d,** Raman Spectra of 6% and 9% Se doped InF$_3$ at an excitation wavelength of 532 nm. No appreciable change in the Raman spectra is observed on varying the amount of Se doping.

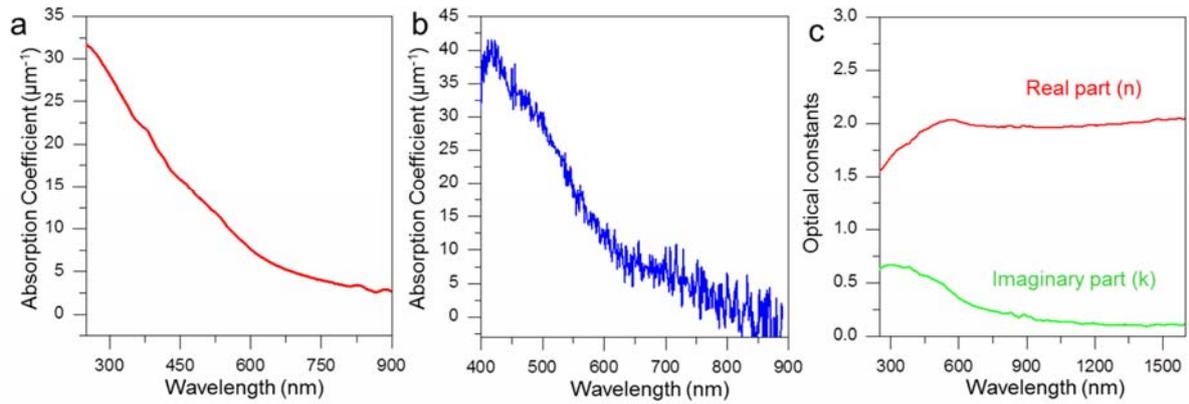

**Figure S6. Optical characterisation of InF$_3$.** Absorption coefficient of **a,** bulk InF$_3$ obtained by the fluorination of bulk InSe and **b,** InF$_3$ obtained by the fluorination of InSe flakes (≈10 nm thick). **c,** Optical constants of InF$_3$ crystal showing real and imaginary parts.

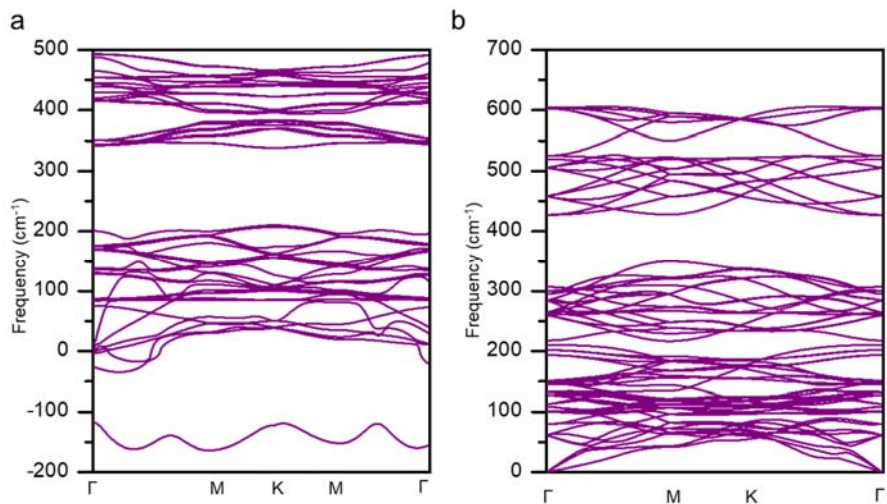

**Figure S7. Phonon dispersion curve for thin InF$_3$ structures.** Phonon dispersion curves for optimised structure of **a**, fully fluorinated bilayer InSe and **b,** fully fluorinated bulk InSe. Our structure optimisation calculation shows that fully fluorinated bilayer InSe is dynamically unstable. This is evident from the corresponding phonon dispersion curves which show a negative frequency throughout the Brillouin zone. Our calculations reveal that fluorination of bi- and monolayers of InSe does not lead to the formation of InF$_3$; however, fluorination of 3 and more layers of InSe leads to a stable InF$_3$ crystal structure. Furthermore, our calculations show that three-layer fluorinated InSe has 6-indium layers with a thickness of ≈1.4 nm, in agreement with the experimental results.

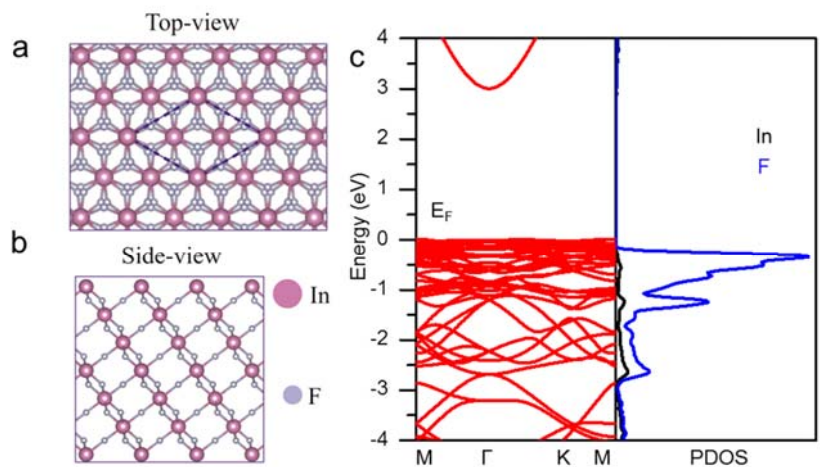

**Figure S8. Bandstructure of InF₃. a,** The crystal structure of pristine InF$_3$. **b**, Side view of the InF$_3$ crystal structure. **c,** Calculated band structure and the corresponding partial density of states (PDOS) of pristine InF$_3$ (colour-coded labels).

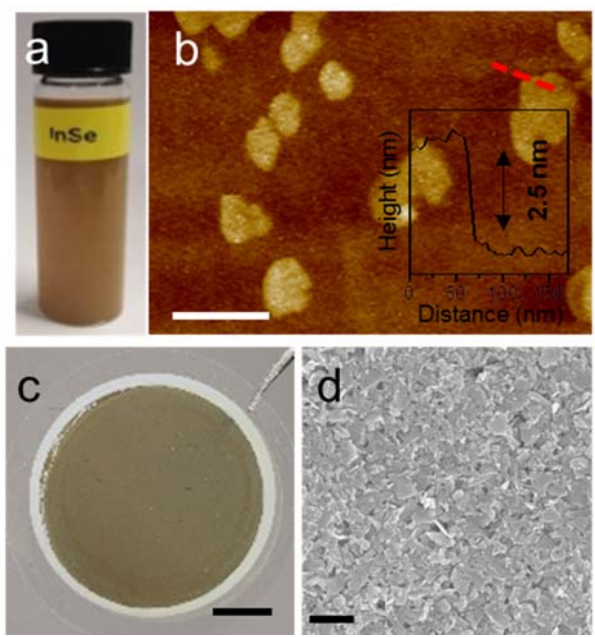

**Figure S9. Liquid exfoliated InSe nanosheets. a,** Photograph of liquid-exfoliated InSe suspension. **b,** AFM image showing few-layer thick liquid exfoliated InSe nanosheets. Scale bar, 200 nm. Height profile along the dashed line is given as inset. Majority of the flakes obtained by liquid exfoliation were found to be < 4 nm in thickness. Samples for AFM were obtained by drop-casting InSe suspension on an oxidised silicon wafer. **c,** Photograph of InSe laminate on Anodisc alumina filter obtained by vacuum filtering the InSe suspension. Scale bar, 5 mm. **d,** SEM image of InSe laminate. Scale bar, 400 nm.

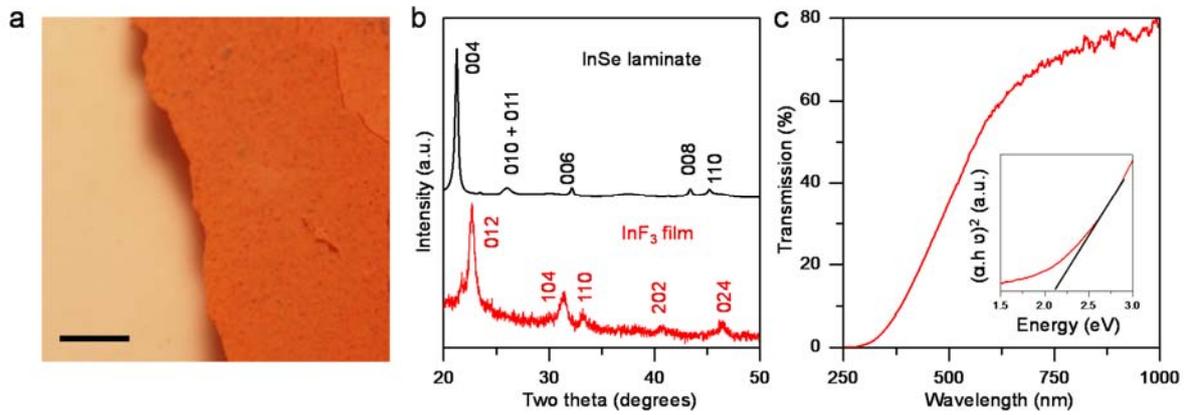

**Figure S10. Large area InF$_3$ film**. **a,** Photograph of a free-standing InF$_3$ film obtained by fluorinating an InSe laminate. Scale bar, 1 mm. **b,** XRD of InSe laminate and fluorinated InSe film. The Miller indices of the Bragg reflections are labelled. **c,** Optical transmission spectra of large area InF$_3$ film. Inset: the associated Tauc plot indicate a direct bandgap of 2.2 eV. Compared to the InF$_3$ crystal, the transition is noted to be broad in InF$_3$ thin films and this could be due to the polycrystallinity or the presence of defects in the sample.